\documentclass[  ,final]  {aipproc}
\layoutstyle{8x11single}
\begin{document}

\title{Phenomenology of the Invisible Universe}
\classification{\texttt{98.80.-k}}
\author{P. J. E. Peebles}{  address={Joseph Henry Laboratories,
 Princeton University, Princeton NJ USA }}

\begin{abstract}
Cosmology is operating now on a well established and tightly constraining empirical basis. The relativistic $\Lambda$CDM hot big bang theory is consistent with all the present tests; it has become the benchmark. But the many open issues in this subject make it reasonable to expect that a more accurate cosmology will have more interesting physics in the invisible sector of the universe, and maybe also in the visible part. 
\end{abstract}

\maketitle

\section{The Cosmological Tests}

\begin{figure}[b]
  \includegraphics[angle=90,height=.33   \textheight]{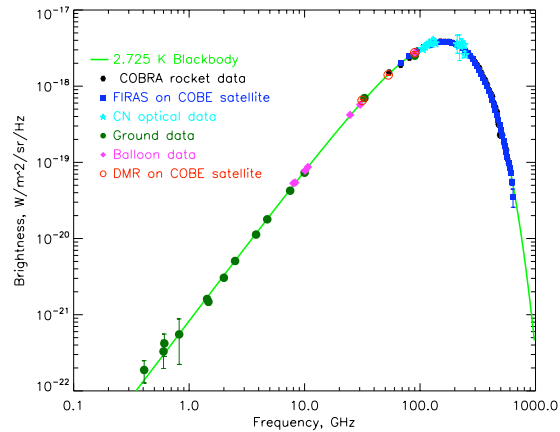}
  \caption{Energy spectrum of the cosmic microwave background, from Alan Kogut.}
\end{figure}

Lest the exciting debates over the great open issues of cosmology cause us to forget, I offer in Figure~1 and Table~1 reminders that we have a firm and substantial empirical basis for our subject. The figure shows the spectrum of the cosmic microwave radiation that nearly uniformly fills space: energy density as a function of frequency. The measurements are diverse and well cross-checked, from the ground, balloons, rocket and satellite missions, and indirectly from spin temperatures derived from interstellar molecular absorption lines. Within the measurement uncertainties, which over much of the range of frequencies in this figure are smaller than the symbols, this wealth of data fits the Planck thermal spectrum at temperature $T=2.725$~K plotted as the solid curve. 

We know the universe as it is now is close to transparent at the frequencies plotted here, because active galaxies at redshifts exceeding unity are detected at these frequencies. Relaxation to this thermal spectrum thus had to have happened earlier in the expansion of the universe, when the density of absorbing matter was great enough make the universe opaque across the Hubble length then. That is, the figure presents us with almost tangible evidence that the universe really did evolve from a very different state. We should take a moment to admire this wonderfully simple and deeply significant phenomenon.

The figure also shows that expansion and cooling to the present transparent state of the universe had to have preserved the thermal radiation spectrum formed when the universe was opaque. Sufficient conditions include standard local physics and a metric description of spacetime. It does not require general relativity theory. It does require that, despite the distinctly inhomogeneous distribution of matter in galaxies and concentrations of galaxies, spacetime is close enough to homogeneous and isotropic not to have produced a sigificant spread of redshifts along different lines of sight (as could have happened in a homogeneous anisotropic universe, for example). Space has to be close to what Einstein envisioned in 1917. 

The $\Lambda$CDM cosmology adopts Einstein's near homogeneous space, his general relativity theory, and his cosmological constant, and it adds the general expansion first discussed by Friedmann and Lema\^\i tre, Tolman's thermal radiation, Gamow's thermonuclear element formation, and the more recent notions of nonbaryonic dark matter and adiabatic near scale-invariant Gaussian initial departures from homogeneity. The application of general relativity is a long extrapolation from its precision tests, but it was natural to try this theory first. Other of these assumptions were not so obvious first guesses; they were adopted because they were seen to offer promising fits to the improving tests. We would be poor theorists indeed if $\Lambda$CDM did not at least approximate what is now measured. But this cosmology has gone beyond that; it is predictive.

\begin{table}[t]
\caption{Cosmological Tests}
\centering
\begin{tabular}{lllll}
\hline
\multicolumn{2}{l}{1. fossil CMB radiation} & 
	\multicolumn{2}{l}{6. cosmic  mass density} \\
& (a) energy spectrum &&  (a) galaxy peculiar velocities \\
& (b) temperature acoustic oscillation &&  (b) gravitational lensing \\
& (c) temperature-polarization cross spectrum & 
	\multicolumn{2}{l}{7. large-scale structure} \\
& (d) integral Sach-Wolfe effect && (a) baryon acoustic oscillation \\
\multicolumn{2}{l} {2. light element abundances} && (b) galaxy N-point functions \\
& (a) deuterium in Ly$\alpha$ absorbers at redshift $\sim 3$ & 
	\multicolumn{2}{l}{8. clusters of galaxies}\\
& (b) helium in dwarf low-metallicity galaxies && (a) mass function as a function of mass \& redshift \\
& (c) observed baryon budget && (b) baryon mass fraction \\
& (d) count of neutrino families & 
	\multicolumn{2}{l}{9. small-scale structure}\\
\multicolumn{2}{l}{3. redshift-magnitude relation} && (a) galaxy structure and evolution \\
\multicolumn{2}{l}{4. stellar evolution \& radioactive decay ages} &&  (b) Lyman-$\alpha$ forest\\
\multicolumn{2}{l}{5. distance scale}\\ 
& (a) distance ladders\\
& (b) gravitational lensing\\
\hline
\end{tabular}
\end{table}

This situation is illustrated in Table~1 (adapted from the book, {\it Finding the Big Bang} \cite{FTBB}; details are there and in other reviews of the state of the cosmological tests). Each measurement named in the table is an independent (or close to it) probe of the universe. Each is capable of falsifying $\Lambda$CDM. This cosmology passes the tests so far. The measurements are difficult, and issues of interpretation and systematic errors are well worth continued close examination. But the diversity of these  probes, and the consistency of independent constraints on the relevant parameters in $\Lambda$CDM, make a close to compelling case that we have not been seriously misled. That certainly does not mean the $\Lambda$CDM cosmology is the final word on the structure and physics of the universe that will be needed to interpret coming generations of measurements. But we can be sure that if $\Lambda$CDM is replaced by something better the new theory will predict a universe that looks much  like $\Lambda$CDM, because that is what is observed. 

\subsection{Precision Cosmology and Accurate Cosmology} 

We should take another moment to admire the abundance of probes of the large-scale nature of the universe represented in Table~1, and to recall the importance of the diversity of these measurements. 

It has been said that we are moving toward precision cosmology, which is correct but misses the point.  A digital scale may read out the weight of an object to many significant figures, in a precise measurement. But if the scale is not well calibrated the measurement may not be very accurate. That is, accuracy is what remains of precision after discounting systematic errors. And our goal is an accurate cosmology. 

In the pursuit of precision cosmology it is natural to concentrate on the relatively few observations that lend themselves to the most precise measurements. This is good science, provided one bears in mind that if the precision measurements are not more numerous than the parameters that may be adjusted to fit them we are getting a cosmology of dubious accuracy. I offer one example, the properties of galaxies. 

The mass distributions in the outer parts of galaxies were an apparent anomaly within the old cosmology, and a hint of something new: dark matter. Recent research proceeds in the opposite direction: take $\Lambda$CDM, dark matter and all, as given, to be used as the basis for analyses of how galaxies acquired their observed  properties. Since galaxies are complicated --- our understanding of star formation and its impact on the diffuse baryons is particularly schematic --- the program must add parameters that are physically well motivated but adjustable because we don't understand how the physics actually is expressed. The freedom of adjustment makes it difficult to judge the significance of the notable advances in this program. The pursuit of more accurate cosmology may be aided by a return to the earlier direction. There are apparent anomalies in the properties of galaxies in terms of what might have been expected in a straightforward reading of numerical simulations of structure formation in $\Lambda$CDM. Instead of asking how the parameters needed to describe the histories of the baryons may be adjusted to remedy this situation consider instead whether apparent anomalies might be hints to better fundamental physics. It is not difficult to invent alternative physics that fits the tests on the scale of the Hubble length about as well as $\Lambda$CDM, but on the scale of galaxies does something different, interesting, and just possibly better. An example of an alternative theory on the scale of galaxies is discussed next; others are in \cite{Khoury} -- \cite{Carroll et al.}.

\section{The Invisible Universe}

\subsection{Dark Matter}

Milgrom's MOdified Newtonian Dynamics, MOND, is a very interesting alternative, an adjustment to gravity physics that allows dark matter to be subdominant to baryons almost everywhere. MOND offers a natural account of the scaling relation $v\propto L^{1/4}$ known as the Tully-Fisher relation between spiral galaxy circular velocity $v$ and luminosity $L$, and the Faber-Jackson relation between stellar velocity dispersion and luminosity of ellipticals. The relation is even tighter when one uses  baryon mass instead of luminosity, and it extends to baryon masses well below what was known when MOND was introduced. That is, on the scales of its intended application MOND is predictive. In these Proceedings Hammer describes observations that indicate spiral galaxies arrive at the $v\propto L^{1/4}$ relation as they form, whether  early or late in the expansion of the universe. This is to be expected in MOND, but curious in $\Lambda$CDM. Ellipticals follow a scaling relation among velocity dispersion, size and luminosity (or baryon mass) that is even tighter than $v\propto L^{1/4}$, and the relation is little affected by environment. I recommend Figure~3 in Bernardi {\it et al.}~\cite{Bernardi} for a striking --- even startling --- illustration of the general insensitivity to what is happening around the galaxy.\footnote{The central generally most luminous galaxy in a rich cluster does systematically differ from the general population, but I am impressed at how small the differences are. For discussions of the evidence and the challenges of its interpretation see \cite{Bern07}.}

How do galaxies ``know'' to join these near universal relations? It is tempting to think the relations might be ``wired into'' the physics, as in MOND. But we know that complex processes operating on smaller scales are capable of producing wonderful scaling relations, and maybe the same is true of galaxy formation. 

MOND is not ruled out by the cosmological tests because it does not make agreed upon predictions on these scales. But MOND faces what I count as a daunting challenge: show how a generalization can fit the tests on larger scales that on the face of it quite systematically point to the dominance of nonbaryonic over baryonic matter. 

I remain  persuaded by the empirical case for nonbaryonic dark matter. But the cautionary example of MOND shows why we should be open to possibly better ideas and watching for phenomena such as the galaxy scaling relations that may be pointing to better physics than the woefully simple dark sector of $\Lambda$CDM. The universe is large, and quite capable of continuing to instruct us. 

\subsection{Dark Energy}

I have not seen any way to avoid the quantum physics prediction of significant contributions to the vacuum energy density. We recall that if the vacuum does not define a preferred frame of motion then its energy appears in the stress-energy tensor in the form of Einstein's $\Lambda$, in elegant agreement with the evidence for $\Lambda$ from cosmology. But, as widely lamented, the value of $\Lambda$ in cosmology seems  absurdly small compared to natural estimates of contributions from quantum physics. Maybe some symmetry forces the net quantum energy to vanish, another arguably natural value. But if so how do we understand the cosmological tests? Is dark energy really present? If so is it the result of a miraculously near but not exact cancellation of quantum contributions, a parameter of the theory -- Einstein's $\Lambda$, or a dynamical entity? If dynamical what is it and how do we understand its apparently curious value? 

\subsubsection{Departures from Homogeneity and the General Expansion of the Universe}

Could the accelerated cosmic expansion usually attributed to $\Lambda$ be instead an effect of the departure from an exactly homogeneous mass distribution? Analyses in these Proceedings illustrate the richly complicated physics of the departure from homogeneity in the classical chaos of matter collecting into galaxies. But I think it is clear that this physics has negligibly small effect on the general expansion. I explain at some length because the issue is important and the physics interesting.

The idea that the general  expansion of the universe is significantly affected by the observed clustering of matter is inspired by analyses of variants of Raychaudhuri's equations. They relate the evolution of the expansion, density, shear and vorticity along the world line attached to  a fluid element. Within the assumption of a pressureless continuous fluid with a differentiable fluid velocity, which is a sensible approximation for this purpose, and apart from the effect of orbit crossings, these relations are identities in general relativity. But deriving the general expansion by averaging the expansions of fluid elements that generally have ended up in the complexity of strongly nonlinear mass concentrations in or around galaxies is an exceedingly challenging problem. 

There is a simpler way \cite{SL} -- \cite{PS}. The mass concentrations in galaxies have characteristic Newtonian gravitational potentials on the order of $\phi\sim v^2\sim 10^{-6}$, where $v\sim 300$ km\,s$^{-1}$ is a typical velocity between galaxies and within galaxies. Since $\phi$ is small the departure from perfectly homogeneous and isotropic spacetime is small and readily analyzed in perturbation theory.  

Consider, for example, that the cosmic microwave radiation has passed through the strongly nonlinear mass concentrations in galaxies. But because the gravitational potentials belonging to these departures from homogeneity are small, and they vary slowly with time, the effect on the radiation is exceedingly small. It is not detected in the radiation energy spectrum shown in Figure~1. The effect is observed \cite{ISW} in the cross-correlation of the variation of the cosmic microwave radiation temperature across the sky (at $\delta T/T\sim 10^{-5}$) with the large-scale galaxy distribution, which is a reasonably useful tracer of mass. This cross-correlation ---  the integrated Sachs-Wolfe effect --- is computed to all needed precision in linear perturbation theory of the departure from a homogeneous and isotropic spacetime produced by the observed clustering of matter.

Computation of the effect of mass clustering on the general expansion of the universe requires some elements of second-order perturbation theory. For simplicity let us model the material content of the universe as a gas of noninteracting dark matter particles with mass $m$ and mean number density $\bar n$. The matter is gravitationally bound in dark halos --- galaxies --- with internal velocities  $v\sim 300$ km~s$^{-1}$ and sizes $r_g\sim 10$\,kpc. Close to equivalent  parameters that measure the effect of the clustering of mass in dark halos on the spacetime geometry are 
\begin{equation}
\epsilon \sim\phi \sim (v/c)^2 \sim r_g/ct\sim 10^{-6},
\end{equation}
where $t$ is the Hubble time. This is a reasonable approximation to our cosmologically flat universe at the present epoch, apart from the subdominant mass in baryons and radiation. Galaxies are nonrelativistic; $\epsilon$ is small. This suggests the line element 
\begin{equation}
ds^2 = a(\tau)^2[(1+2\phi)d\tau^2 - (1 - 2\phi)\delta _{ij}dx^idx^j],
\label{eq:1}
\end{equation}
which reduces to the nonrelativistic Newtonian form on the scale of galaxies, and describes the general expansion of the universe in the average over much larger scales. 

In this line element the time-time part of Einstein's field equation to zeroth order in $\epsilon$ is
\begin{equation}
R_{00} = -3{d\over d\tau}{1\over a}{da\over d\tau} + \nabla^2\phi =
4\pi Gm\bar na^2(1+\delta),
\end{equation}
where $\delta(\vec x,t) = n(\vec x,t)/\bar n(t) - 1$ is the dimensionless measure of the departure from a homogeneous mass distribution. In the central parts of galaxies $\delta$ is much greater than unity, but that does not concern us; we are interested in $\phi$, which we know is small (apart from neutron stars and black holes, an interpretation of which is offered below). The standard and sensible practice is to separate this relation into the two parts
\begin{equation}
{\ddot a\over a} = - {4\over 3}\pi Gm\bar n,\qquad  \nabla^2\phi = 4\pi Gm\bar na^2\delta,
\label{eq:delsqphi}
\end{equation}
where the dots are derivatives with respect to proper time ($dt = ad\tau$), and the boundary condition is that $\phi$ has no source-free part. This means the spatial average of $\phi$ in the coordinates  of Eq.~\ref{eq:1} vanishes along with the spatial average of $\delta$. 

To find the effect of mass clustering on the Friedmann-Lama\^\i tre equations to order $\epsilon$ we must compute spatial averages of Einstein's field equation to order $\phi^2$ for the dominant term, 
\begin{equation}
t^2\langle\nabla\phi\cdot\nabla\phi\rangle \sim\epsilon.
\label{eq:nonlinearpart}
\end{equation}
To order $\epsilon$ we can evaluate Eq.~\ref {eq:nonlinearpart} using the zeroth order equation for $\phi$ in Eq.~\ref{eq:delsqphi}. It is an interesting exercise to complete the calculation (which requires modest extensions of the computations in  \cite{SL} -- \cite{PS}) and check that the spatial averages of the time and space parts of Einstein's equation yield the usual two Friedmann-Lema\^\i tre equations in which the effective mean mass density and pressure terms are 
\begin{equation}
\bar\rho = m\bar n (1 + K + W), \quad \bar p = m \bar n (2K + W)/3, \quad K= \langle v^2\rangle/2, \quad W = -{1\over 2}Gm\bar n a^2\int{d^3x\over x}\langle\delta(\bf r)\delta(\bf x - \bf r\rangle).
\label{eq:eff_pressure}
\end{equation}
These are in the coordinates of Eq.~\ref{eq:1}, which to order $\epsilon$ are what an astronomer uses to measure $\delta$ and $v$. The mean kinetic energy per unit mass, $K$, comes from the stress-energy tensor for the collisionless particle model. The gravitational energy per unit mass, $W$, has contributions from the stress-energy tensor and from the term proportional to Eq.~\ref{eq:nonlinearpart} in the left-hand side of the field equation. 

For an intuitive understanding of Eq.~\ref{eq:eff_pressure} consider the stress-energy tensor of a galaxy. We know from the nonrelativistic kinetic theory of gasses that in the galaxy rest frame an isotropic dark matter particle velocity dispersion has pressure $m\bar n\langle v^2\rangle/3$, which appears in the diagonal space part of the stress-energy tensor. It is not surprising that the space average of this expression is the term $2m\bar nK/3= m\bar n\langle v^2\rangle/3$ in Eq.~\ref{eq:eff_pressure}. Now consider the spatial part of the total stress-energy tensor. In a statistically isotropic solid with a free boundary the space parts of the stress-energy tensor have zero mean. That comes about because the interaction holding the solid together adds to the stress-energy tensor tensor a term that just cancels the contribution by the velocity dispersion of the ions. In a gravitationally bound and stationary galaxy the virial theorem says the gravitational energy per unit mass, $W$,  and the kinetic energy per unit mass, $K$, satisfy $W = -2K$, meaning $\bar p$ vanishes in Eq.~\ref{eq:eff_pressure}, in analogy to the solid. A non-rotating neutron star has a large internal pressure, again just balanced by its internal gravity: the galaxy sees it as a particle, to excellent accuracy. The orbits of stars around the massive black hole in the center of a galaxy with a well-developed bulge show that the stars  see the black hole as an ordinary point mass, to excellent accuracy, and it is scarcely surprising that the rest of the universe sees it in the same way. The gravity holding a galaxy together sees a considerable mass deficit in a neutron star or black hole. And, as expected, the effective mass density $\bar\rho$ in Eq.~\ref{eq:eff_pressure} is lower than the rest mass density $m\bar n$ by the amount of the binding energy per unit mass, $-(K+W)$. Finally, I offer another interesting exercise: following \cite{SF} check consistency of Eq.~\ref{eq:eff_pressure} with the relation between $K$ and $W$ that Irvine \cite{Irvine} derived in 1961, 
\begin{equation}
{d\over dt}(K + W) =-{1\over a}{da\over dt}(2K+W).
\end{equation}
To avoid confusion I repeat that this paragraph is not a derivation of Eq.~\ref{eq:eff_pressure} but rather an explanation of why I find the result persuasive. 

The large-scale growing mass fluctuations in the real universe are not in statistical equilibrium, so $2K + W$ does not vanish. These fluctuations contribute to $\bar p$, in an analog of sorts to the integrated Sachs-Wolfe effect. Also, Eq.~\ref{eq:eff_pressure} is a spatial average, which is appropriate for the effect of spacetime variations along a line of sight but not for the potential $\phi$ at the observer. But these corrections to the Friedmann-Lema\^\i tre equations are on the order of $\epsilon$, roughly a part in $10^{6}$, well below what would needed to vitiate the evidence for detection of $\Lambda$. I respect the deep analyses that start from Raychaudhuri's equations, but I think the situation is clear. Within general relativity theory we must learn to live with $\Lambda$.

\subsubsection{Modified Gravity}

One can avoid that, eliminate $\Lambda$, by adjusting the gravity physics, as also is  well discussed in these Proceedings. One approach is to choose a Lagrangian density for gravity operating four-dimensional spacetime that is close enough to the Ricci scalar $R$ of general relativity theory in regions of high density and on relatively small scales to pass the precision tests of gravity physics, and on large scales acts as a close enough surrogate for $R$ plus a constant to pass the cosmological tests. One can choose a function $f(R)$ that fits these conditions and passes the present tests of gravity and cosmology about as well as general relativity with a cosmological constant --- provided quantum physics has no net effect on the vacuum. If more precise and accurate tests prove to be  inconsistent  with $R$ plus a constant and agree with an $f(R)$ theory it will be a really profound advance. 

Other ways to modify gravity are discussed in these Proceedings. I depart from my charge, the phenomenology of the invisible universe, to offer my supposition that the provenance within some deeper unified theory of everything of a successful modified gravity would be an even greater puzzle than the standard $R$  plus a constant. I expect the community by and large will continue to work with $R$ plus a constant in $\Lambda$CDM until driven from it. And  it is right and proper that others will continue to prepare the way for the eventuality that something better than $R$ plus a constant proves to be needed. 

\subsubsection{Learning to Live with $\Lambda$}

In the $\Lambda$CDM cosmology what might be the nature of this mysterious dark energy component, and what might we make of its curious value?

The evidence is that our universe now is becoming dominated by dark energy, in effect approaching another epoch of inflation but at really low energy. It is natural to consider the possibility that the dark energy manifest in the cosmological tests is a remnant of the dark energy that drove inflation; maybe the present dark energy density is small because it has been rolling toward zero for a long time. These Proceedings document  the rich considerations in research on the physics of inflation that may also shed light on the nature of the present  Invisible Universe. 

A demonstration that the dark energy density actually is evolving, as opposed to Einstein's constant $\Lambda$, certainly would stimulate thinking about the nature of dark energy, though what the community would think about is less clear; we are short of manifestly promising ideas. A detection of variation of the strength of the electromagnetic interaction, or of other of the parameters of physics, also would be a deeply influential advance. The same would be true of detection of H{\small I} radiation (redshifted 21-cm emission from atomic hydrogen) from reionization, or of a characterization of the dark matter, or of a third generation all-sky position-magnitude-redshift survey, or of a really searching and complete study of how stars form as a function of environment. Research aimed at the first of these problems, detecting or better constraining evolution of the dark energy density, certainly merits community support. But I don't think our subject is ready for a dark energy mission in the big science mode of the search for the Higgs boson; there is still too much other work to be done in straightening the foundations of our subject.

What are we to make of the value of $\Lambda$? Most widely discussed is the anthropic argument, as well framed by Weinberg \cite{Weinberg}: among the ensemble of all possible universes, we have to have found ourselves in one that allows the formation of galaxies of stars that last some $10^{10}$ years. This is an elegant answer, but at the expense of a new question: how do we determine whether the anthropic argument is more than a ``just so'' story invented to save the phenomenon? 

On the positive side, the anthropic argument offers a possible  window into the deeper physics that allows many universes and, in universes in which dark energy is a meaningful concept, some hint to the range of  values the deeper physics allows this parameter. It is noteworthy that, at a time when $\Lambda$ was not at all popular,  Weinberg's considerations along these lines led him to conclude that the dark energy density might be expected to be larger than the present matter density, which has proved to be the case. Similar arguments have been applied to other of the many parameters in our physics that are so very well and finely tuned to allow our existence, from the strength of the strong interaction to the expansion of water on freezing. This offers fascinating lines of investigation, but I would be more comfortable with a less anthropocentric way to put it, as follows.

In another universe different physics may allow development of systems that become so deeply complex as to ``wonder why'' they are in a universe that is so wonderfully well suited to the requirements of their existence. (The phrase, ``wonder why,'' is anthropocentric, of course, but I don't have a better way to put it.) The answer many of us would be happy with is that that is where these systems evolved. Maybe somewhat fewer would be happy with the similar argument that all the details of all the physics of our universe are exquisitely well suited for our existence because this is where we evolved, but it seems logical to me. In this way of putting it the  challenge is to  determine the bounds on the values of $\Lambda$ and on all the rest of physics in the class of all possible universes in which systems can get complex enough to start ``wondering why'' conditions are what  they are. 

\section{Summary Remarks}

Cosmology has changed from the mode of operation the older of us remember without nostalgia. We are now tightly constrained by a rich observational basis that makes it exceedingly delicate --- though certainly not impossible --- to find and demonstrate viability of alternatives to the $\Lambda$CDM cosmology. We each have chosen a mode of research appropriate for this relatively new situation in cosmology. It is worth listing strategies, to remind ourselves and funding agencies that we do have choices.

It is entirely reasonable to take the conservative position that $\Lambda$CDM, maybe with slow evolution of the dark energy density and fine tuning of initial conditions, includes all the physics that is going to be relevant to analyses of phenomenological cosmology. Then the goal becomes to understand how this given physics is expressed in the processes that make the universe we observe, from the Hubble length down to galaxies and stars and planets, and back in time to light element production and maybe baryogenesis. This strategy has proven its worth by defining a host of productive research problems. 

If $\Lambda$CDM does differ from reality enough to matter it means this conservative strategy will reveal it  through identification of phenomena that cannot be reconciled with the cosmology. But beware of precision cosmology, because the observations that lend themselves precision measurements reduce the tests in Table~1 to a number perilously close to the parameters we are willing to adjust in $\Lambda$CDM. (Currently discussed additions to the more traditional cosmological parameters include evolution and maybe spatial gradients of the dark energy density, a soup\c{c}on of annihilating or collisional or warm dark matter, or maybe  even a new long-range interaction in the dark sector. A competent theorist likely could offer quite a few other ways to save the phenomena within $\Lambda$CDM.) I expect that the better strategy for finding a more accurate cosmology will be to examine all measurements, however inaccurate, that probe different aspects of the universe and may challenge standard ideas. 

A more proactive strategy for discovering a possibly more accurate cosmology is to search for apparent anomalies in $\Lambda$CDM, situations that seem particularly challenging for this theory. My list commences with a pronounced difference between the physics of the visible and invisible universes. Physics in the visible sector is wonderfully simple, in the appropriate sense, but its expression is spectacularly complicated. Physics in the invisible sector is so exceedingly simple that its expression is simple: the dark matter just piles up in halos and the dark energy sits there or maybe slowly evolves. It is wishful thinking, but also sensible, to suspect that this apparently anomalous situation is telling us that we have settled for the simplest physics we can get away with at the present crude levels of probes of this sector. It presents no hints of phenomena to look for; I read it an invitation to look broadly. 

Other apparent anomalies may offer more direct hints of better physics. Examples include preliminary results from underground dark matter particle detectors and the search for signatures of dark matter annihilation. At the time of writing the experimental/observational situation is confused, maybe in part a result of systematic errors in difficult measurements, maybe in part because we are seeing hints to something new. Other examples are the more curious properties of galaxies.  How did galaxies arrive at their scaling relations that show such little regard for environment? Why do the voids defined by normal galaxies look so very empty? Why do most of the nearest spiral galaxies edge up to the very empty Local Void? How did massive black holes in active galactic nuclei form so early in the expansion of the universe? Why is large-scale structure in the galaxy distribution so very large?  If such curiosities are expressions of standard physics then understanding how they came about will teach us something of value. And maybe some are hints to better physics, which would be a good thing too. 

Research in the proactive direction may accept the standard physics of the visible universe and seek something more interesting in the poorly explored invisible part. As one sees in these Proceedings, there is no shortage of ideas here. The adoption of standard visible sector physics is conservative, but maybe overly so. For example, the precision tests of general relativity are on scales some fifteen orders of magnitude smaller than the Hubble length. It is sensible to continue to question this enormous extrapolation, to consider modifications of gravity physics. This strategy has led to viable alternatives that to date seem to me to be less logically compelling than general relativity with $\Lambda$CDM. This is not a very convincing argument, but I expect it will lead most in the community to continue to work within the physics of $\Lambda$CDM unless/until some new phenomenology or really attractive idea drives us from it. 

I refer to Wigner's \cite{Wigner} essay, {\it The Unreasonable Effectiveness of Mathematics in the Natural Sciences},  for the arguments that lead me to expect that the community will not arrive at the one true cosmology, if such a thing exists. We will instead end up with the best approximation allowed by the limited ability to observe. We cannot rule out the possibility that something will turn up that invalidates our present cosmology, but from all experience that is exceedingly unlikely. Much easier to imagine is fine tuning to fit new measurements will lead to a more accurate cosmology that contains elements of $\Lambda$CDM along with new and interesting adjustments.


\begin{thebibliography}{9}

\bibitem{FTBB} P. J. E Peebles, L. A. Page and R. B. Partridge, \emph{Finding the Big Bang}, Cambridge University Press, Cambridge, 2009.

\bibitem{Khoury} N. Afshordi, G. Geshnizjani and J. Khoury, \emph{Journal of Cosmology and Astro-Particle Physics} \textbf{8}, 30 (2009).

\bibitem{Keselman et al.} J.A. Keselman, A. Nusser and P.J.E. Peebles,  arXiv:0902.3452 (2009).

\bibitem{Carroll et al.} S.M. Carroll, S. Mantry, M.J. Ramsey-Musolf and C.W. Stubbs, \emph{Physical Review Letters} \textbf{103}, 011301 (2009).

\bibitem{Bernardi} M. Bernardi, R.C. Nichol, R.K. Sheth, C.J. Miller and J. Brinkmann, \emph{Astronomical Journal} \textbf{ 131}, 1288 (2006).

\bibitem{Bern07} M. Bernardi, J.B. Hyde, R.K. Sheth, C.J. Miller and R.C. Nichol, R.~C. \emph{Astronomical Journal} \textbf{133}, 1741 (2007).

\bibitem{SL} U. Seljak and L. Hui, in \emph{Clusters, Lensing,  and the Future of the Universe}, edited by V. Trimble and A. Reisenegger, Astronomical Society of the Pacific, San Francisco, 1996, p. 267.

\bibitem{SF} E.~R. Siegel and J. N. Fry, \emph{Astrophysical Journal Letters} \textbf{ 268}, L1 (2005).

\bibitem{KAF} M. Kasai, H. Asada and T. Futamase, \emph{Progress of Theoretical Physics}  \textbf{115}, 827 (2006).

\bibitem{PS} A. Paranjape and T.P. Singh, \emph{Physical Review Letters} \textbf{101}, 181101(2008).

\bibitem{ISW} T. Giannantonio, R. Scranton, R.G. Crittenden, R.C. Nichol, S.P. Boughn, A.D. Myers and G.T. Richards, \emph{Physical Review D} \textbf{77} 123520 (2008).

\bibitem{Irvine} Irvine, W.~M. Ph.D.~Thesis, Harvard University (1961).

\bibitem{Weinberg} S. Weinberg, \emph{Physical Review Letters} \textbf{59}, 2607 (1987).

\bibitem{Wigner} E. P. Wigner, \emph{Communications in Pure and Applied Mathematics} \textbf{13}, 1 (1960),

\end{thebibliography}
\end{document}